\documentclass{aa}  
\usepackage{booktabs} 
\usepackage{amsmath} 
\usepackage{graphicx}
\usepackage{amssymb}
\usepackage{float}
\usepackage{txfonts}
\usepackage{gensymb}
\usepackage{hyperref}
\usepackage{subfig}
\usepackage{threeparttable}
\usepackage{multicol}
\usepackage{multirow}
\usepackage{longtable}
\usepackage{array}
\usepackage{caption}
\usepackage{rotating}
\usepackage{tabularx}
\usepackage{afterpage}
\usepackage{placeins}
\usepackage[normalem]{ulem}
\usepackage[dvipsnames]{xcolor}
\usepackage[mathlines,switch]{lineno}

\newcommand\masyr{\ensuremath{\text{mas$\cdot$yr}^{-1}}}

\newcommand\kms{\ensuremath{\text{km$\cdot$s}^{-1}}}

\newcommand\pmra{\ensuremath{\text{$\mu_{\alpha}^*$}}}
\newcommand\pmdec{\ensuremath{\text{$\mu_{\delta}$}}}

\begin{document} 

\authorrunning{Hu et al. 2026}

\titlerunning{Missing pairs in open cluster catalogs}

\title{Missing pairs in open cluster catalogs}

 \author{Qingshun Hu \inst{1, 2}, Yufei Cai \inst{1}, Caroline Soubiran \inst{2}, Yu Dai \inst{1}, Yuting Li \inst{1}, Yangping Luo \inst{1}, and Mingfeng Qin\inst{3,4}}
	
\institute{School of Physics and Astronomy, China West Normal University, No. 1 Shida Road, Nanchong 637002, People's Republic of China, (\email{qingshun0801@163.com}\label{inst1}) \and Laboratoire d'Astrophysique de Bordeaux, Univ. Bordeaux, CNRS, B18N, all\'ee Geoffroy Saint-Hilaire, 33615 Pessac, France, (\email{caroline.soubiran@u-bordeaux.fr}\label{inst2}) \and Institute for Frontiers in Astronomy and Astrophysics, Beijing Normal University, Beijing 102206, People's Republic of China \label{inst3} \and School of Physics and Astronomy, Beijing Normal University, Beijing 100875, People's Republic of China \label{inst4}}

\date{Received 27 March, 2026; Accepted 16 April, 2026}

\abstract
 {Open clusters (OCs) in our Galaxy can be found in pairs, possibly forming physical binaries, or in groups. These objects offer unique insights into the process of star formation and testify to the dynamical interactions at local and galactic scales. Therefore, building as complete a census as possible is  a valuable endeavor.}
 {This work is aimed at identifying and characterizing new OC pair candidates that had been overlooked in previous studies.}
 {Two recent comprehensive catalogs were cross-matched to identify OCs in the first catalog that had been missing from the second one. From this list, counterparts in the second catalog were searched within a 3D distance of 50~pc. Candidate pairs were then selected by applying constraints on the tangential velocity (TV) difference. An orbital integration was performed to assess gravitational binding. The similarity in terms of the radial velocity (RV) and age was evaluated.}
 {We identified seven isolated binary cluster candidates, comprising two likely bound systems with stable orbits over 100 Myr; two pairs with a possible common origin but lacking RV confirmation; and three pairs with significant velocity discrepancies, suggesting they are unbound or in transitional states. We also identified six cluster group candidates, while refining the membership of known complexes such as UBC\_672 and NGC\_1977, and discovering a new group around FSR\_0198. Notably, the UBC\_392 group exhibits coherent proper motions but inconsistent RVs and large age spreads, indicating that it is not gravitationally bound. Additionally, we reconciled 15 clusters with discrepant nomenclature between the two catalogs.}
{Multi-catalog integration combined with kinematic and dynamical validation is essential for establishing a complete census of Galactic cluster pairs. Our findings have effectively expanded the known binary cluster sample and provided refined targets for future studies.}


\maketitle
\nolinenumbers

\section{Introduction}

Open clusters (OCs) in the Milky Way are occasionally found in pairs or higher order multiple systems \citep{Subramaniam1995, Vazquez2010, Song2022}. The origin of such systems may be primordial, arising when several clusters form nearly simultaneously or sequentially through the fragmentation of a single molecular cloud \citep{delaFuenteMarcos2009a}. In this scenario, we would expect the clusters to be spatially close and to exhibit comparable ages, kinematics, and chemical compositions. These systems are of particular interest, as they provide key insights into the star-formation activity within molecular clouds as well as into the mechanisms and timescales of tidal disruption. Alternatively, cluster pairs may form through tidal capture during close encounters between unrelated clusters \citep{deOliveira2002, Grinenko2025}, or through resonant trapping. The subsequent dynamical evolution of such pairs depends on processes operating both locally and on Galactic scales. Apparent cluster pairs may also arise as optical systems, when unrelated OCs overlap in projection along the same line of sight.

The advent of the Gaia mission \citep{GaiaCollaboration2016} has significantly advanced both the census and the characterization of OCs, as recently reviewed by \citet{Cantat-Gaudin2024}. The high-precision, all-sky astrometric and photometric data provided by the Gaia catalogs have made the identification and study of gravitationally bound OC systems more efficient and reliable. Investigating the formation and evolution of such systems through the determination of their physical properties has become a particularly active field of research. Nevertheless, the true fraction of OCs in the Milky Way that belong to binary or multiple systems remains an open question.

Since the release of Gaia data in recent years, an increasing number of double or binary clusters \citep{Liu2019, Casado2021, Song2022, Qin2023, Li2025, Palma2025, Liu2025, Hu2025} have been identified across various star cluster catalogs \citep{Cantat-Gaudin2020, vanGroeningen2023, Hunt2024}. In particular, \citet{Palma2025} published the most extensive census of binary cluster candidates to date, with 617 pairs and 261 groups, based on their analysis of the cluster catalog of \cite{Hunt2024}. \citet{Liu2025} also provided a binary cluster catalog (400 pairs) based on \cite{Hunt2024}.  However, these searches have almost exclusively relied on individual catalogs, potentially overlooking some double or binary clusters whose constituent clusters do not appear together in the same catalog.

In this work, we report new candidates of OC pairs and groups. They were found by comparing the catalog of OCs by \citet[][CG20 hereafter]{Cantat-Gaudin2020}, based on Gaia Data Release 2 \citep[DR2,][]{GaiaCollaboration2018} and that of \citet[][H\&R24 hereafter]{Hunt2024} based on Gaia Data Release 3 \citep[DR3,][]{GaiaCollaboration2023}.

The paper is organized as follows. We describe the data sources adopted and the processes to identify cluster pair candidates in Section~\ref{data} and Section~\ref{method}, respectively. In Section~\ref{the_same_clusters}, we present a list of clusters which are the same ones but with different names between CG20 and H\&R24. Sections~\ref{pairs} and \ref{group} focus on the cluster pair and group candidates, respectively. We summarize our results and conclusions in Section~\ref{summary}.

\section{Open cluster catalogs}\label{data}

Prior to the Gaia era, several large catalogs had compiled over 1,000 star clusters \citep{Dias2002, Kharchenko2013}, although none of them contained more than 3,010. Owing to differing detection methods and input data, these catalogs often varied significantly from one to another. Following the release of Gaia data, new large-scale censuses emerged in rapid succession, expanding the known population from just over 2,000 to more than 8,000 clusters \citep{Bica2019, Kounkel2020, Cantat-Gaudin2020, Hunt2023}. Despite their common foundation on Gaia data, the number of clusters detected varies across these surveys, as each employs distinct methodologies and selection criteria. As an effort to build a complete, up-to-date list of all known star clusters, \citet{Perren2023} compiled the Unified Cluster Catalog\footnote{\url{https://ucc.ar/}} (UCC), which cross-matches over 30 different OC catalogs and re-processes all clusters using their own membership algorithm on Gaia DR3 data, consolidating cluster nomenclature across multiple surveys.

For this work, we used the two large catalogs of star clusters and their members published by CG20 and H\&R24. The former catalog comprises 2,017 clusters, including $\sim$230,000 members, identified through the unsupervised classification scheme UPMASK \citep{Krone-Martins2014} applied to Gaia DR2. It includes all of 1,481 clusters from \citet{Cantat-GaudinAnders2020}, the firstly discovered 550 {\it UBC} clusters discovered by \citep{Castro-Ginard2018, Castro-Ginard2019, Castro-Ginard2020}, and 35 {\it LP} clusters discovered by \citet{Liu2019}. The H\&R24 catalog is three times larger, totaling 7,167 star clusters, with more than 1 million members, including OCs, globular clusters, and moving groups, obtained through the Hierarchical Density-Based Spatial Clustering of Applications with Noise (HDBSCAN) \citep{McInnes2017} applied to Gaia DR3. \citet{Hunt2023} report that 96.6\% of the clusters from \citet{Cantat-GaudinAnders2020} could be retrieved by this method, this ratio being 89.2\% and 75.0\% for the {\it UBC} clusters from \citet{Castro-Ginard2020} and the {\it LP} clusters from \citet{Liu2019}, respectively. Since none of these three detected ratios reached 100\%, it means that some clusters in CG20 must be absent from H\&R24. Moreover \citet{Hunt2023} have not compared their list to the full catalog of CG20. In this paper, we investigate the clusters from CG20 that are missing in H\&R24, but close to another OC, as described below.

\section{Identification of cluster pair candidates}\label{method}

We first cross-matched the CG20 and H\&R24 catalogs based on cluster names, identifying a total of 1,904 clusters in common. The matching was performed using the main cluster name listed in the \texttt{Name} column of H\&R24, and the other identifiers listed in the \texttt{AllNames} column. We then focused on the 113 remaining clusters from CG20 that appeared to be missing in H\&R24.

In order to ensure uniformity between the two catalogs, we updated the member data of the 113 clusters from CG20, originally based solely on Gaia DR2, by retrieving their Gaia DR3 values. For each cluster from CG20, its age was taken directly from CG20, while the other parameters (right ascension, $\alpha$; declination, $\delta$; parallax, $\omega$; proper motions, $\pmra$ and $\pmdec$; and radial velocity, RV) were re-estimated using Gaia DR3 data, with values and uncertainties given by the median and median absolute deviation (MAD) of the members. The parameters for H\&R24 clusters were adopted directly from H\&R24.

Our approach to determining possible cluster pairs was to identify, for each of these 113 OCs, the counterparts in the H\&R24 catalog lying within 50 pc, which is the distance criterion adopted by \citet{Palma2025} to search for binary clusters in H\&R24. Thus, we were able to compute the 3D distances ($\Delta$3D) between the 113 missing clusters and the 7,035 clusters from H\&R24, corresponding to their full list where 132 globular clusters were removed. We found 101 cluster pairs with $\Delta$3D~$\leq$~50~pc, corresponding to 30 distinct CG20 clusters each paired with one or more H\&R24 neighbors. The next step was to investigate whether the TVs, and RVs (when available) of both components in these 101 pairs might be compatible. Ages from the original catalogs were also compared. As noted by \citet{delaFuenteMarcos2009a, Palma2025}, close spatial proximity without shared kinematics and comparable age is likely indicative of a chance alignment, rather than a bound system.

\begin{figure*}[htbp]
   \centering
    \includegraphics[width=145mm]{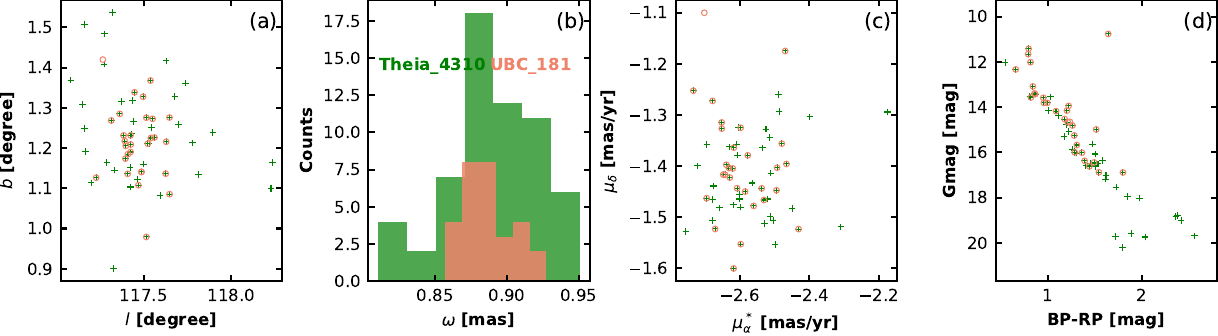}
      \caption{Distributions of position(a), parallax (b), proper motion (c), and CMD (d) of clusters being the same object with different names in CG20 (pink) and H\&R24 (green). }
       \label{fig:the_same_cluster}
\end{figure*}

The TV of each cluster cluster was estimated using the formula 4.74~$\times$~$\mu$~$\times$~$\omega^{-1}$, where $\mu$ is the total proper motion in \texttt{mas/yr} and $\omega$ is the parallax in \texttt{mas}. The median and MAD of the TVs were derived from 100 Monte Carlo (MC) samples, propagating the uncertainties in $\pmra$, $\pmdec$, and $\omega$, along with their dispersions given in Table~\ref{tab:pairs_para}. The direction of the transverse motion, namely, the position angle, $\theta$, between the direction of celestial north and that of the proper motion of clusters (measured eastward from the north) was similarly calculated through 100 MC samples.

Among the 101 possible pairs, we found a few of them to be the same clusters with different names in CG20 and H\&R24 (see next section). We then identified pairs with $\Delta$TV~<~10~\kms and $\Delta$$\theta$~<~20$^{\degree}$ that are binary cluster candidates, with some of them belonging to the same group. We chose those values by analyzing the typical difference in transverse motion between the components of well-established pairs from the literature \citep[e.g.][]{Conrad2017, Liu2025}. We present our final seven binary cluster candidates in Sect.~\ref{pairs} and six cluster group candidates in Sect.~\ref{group}. These cluster pairs have not been reported in previous studies \citep[e.g.][]{Palma2025, Liu2025}. The groups are already known, but we provide new components to them. These findings are discussed in the following sections.

The remaining 83 objects from CG20 with no counterpart are real OCs missing in H\&R24. We provide their astrometric parameters updated with Gaia DR3 in the accompaying online table.

\begin{table}[htbp]\tiny
\centering
\caption{Same clusters with different names.}
\renewcommand{\arraystretch}{1.5}
\resizebox{\columnwidth}{!}{
\begin{tabular}{lclccc}
\hline \hline 
Cluster$_{\text{CG20}}$ & Number$_{\text{CG20}}$ & Cluster$_{\text{H\&R24}}$ & Number$_{\text{H\&R24}}$ & In common & Identified  \\ 
\hline
UBC\_181 & 29 & Theia\_4310 & 60 &  28  & y \\
UBC\_191 & 95 & FSR\_0596 & 27 & 21  & y \\
UBC\_480 & 13 & Alessi\_34 &  857 &  13  & y \\
UBC\_14 & 45 & CWNU\_1074 &  217 &  35   & n \\
UBC\_526 & 17 & HSC\_2576 & 198 & 17  & y \\
NGC\_2645 & 34 & Pismis\_6 & 136 & 26   & y \\
UPK\_533 & 55 & UPK\_545 & 597 & 48 & n \\
UBC\_55 & 56 & FSR\_0686 & 109 & 47 & y \\
UBC\_392 & 33 & UPK\_194 & 124 & 33  & y \\
UBC\_354 & 21 & LP\_2117 & 883 & 21 & y \\
UBC\_73 & 49 & Gulliver\_56 & 139 & 43 & y \\
UBC\_382 & 24 & UBC\_591 & 78 & 22 & y \\
UBC\_450 & 38 & UBC\_634 & 363 & 34 & y \\
UBC\_370 & 20 & UBC\_583 & 73 & 13 & n \\
UBC\_186 & 522 & UBC\_602 & 108 & 67 & n \\
\hline
\end{tabular}
}
\tablefoot{Cluster names in CG20 and H\&R24 with the corresponding number of members in each catalog and the number of common members. Those objects identified by \citet{Perren2023} listed in the "identified'' column are marked by `y', while `n' indicates those detected by this work for the first time.}

\label{tab: The_same_clusters}
\end{table}

\section{Same clusters with different names}
\label{the_same_clusters}

H\&R24 identified clusters listed under different names in earlier catalogs. From the 101 possible pairs mentioned above, we found 15 correspondences between CG20 and H\&R24 that appear to represent the same physical clusters, as presented in Table~\ref{tab: The_same_clusters}. For instance, although the \texttt{AllNames} entry for cluster \texttt{Theia\_4310} in H\&R24 does not include the designation \texttt{UBC\_181} used in CG20, we confirm they are the same object, as shown in Fig.~\ref{fig:the_same_cluster}. This is established through their overlapping distribution in position, proper motion, parallax, and color-magnitude diagram (CMD), along with the identification of 28 common members. The same methodology confirms 14 other pairs. These identifications consolidate cluster nomenclature across different catalogs.

The issue of duplicate clusters is also addressed by 
\citet{Perren2023} in the UCC, which compiles over 18,000 clusters from the literature. The UCC
employs a parallax-based decision rule to flag probable duplicates, 
comparing the 5D distances (coordinates, parallax, and proper motions) 
between clusters and converting them into duplication probabilities. 
Notably, the UCC re-processed all clusters using their own membership 
algorithm on Gaia DR3 data, rather than relying on published member lists 
from the original catalogs. This approach may lead to different conclusions when the original member assignments in CG20 or H\&R24 differ significantly from the UCC's re-derived membership. The UCC is designed to be a living catalog that undergoes periodic updates as new cluster candidates are published, having grown from approximately 14,000 objects at its initial release to over 18,000 at present. However, only about 7500 objects among them are considered to be reliable OCs.

We compared our list of duplicate clusters with the UCC database. As shown in Table~\ref{tab: The_same_clusters}, our independently identified duplicate pairs are generally consistent with the UCC results. For example, our analysis shows that UBC\_181 and Theia\_4310 
share 28 common members, consistent with the UCC which flags 
these two clusters as sharing 70.2\% of their members. However, four pairs in our list (UBC\_14 / CWNU\_1074, UPK\_533 / UPK\_545, UBC\_370 / UBC\_583, and UBC\_186 / UBC\_602) are not flagged as duplicates in the UCC, likely due to differences in the membership determination methods employed. These discrepancies highlight the importance of cross-validating duplicate identifications across different methodologies.

\section{Binary cluster candidates}\label{pairs}

\begin{figure*}[htbp]
   \centering
   \includegraphics[width=160mm]{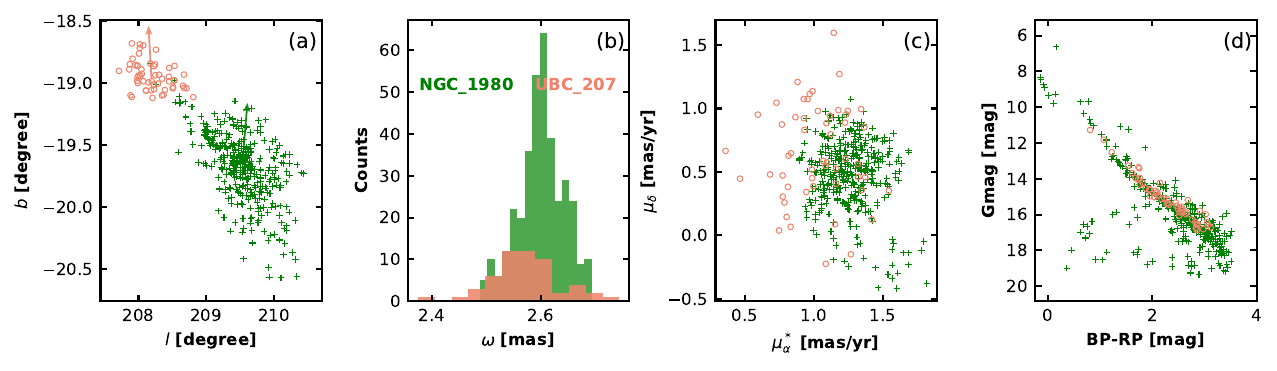}
    \includegraphics[width=160mm]{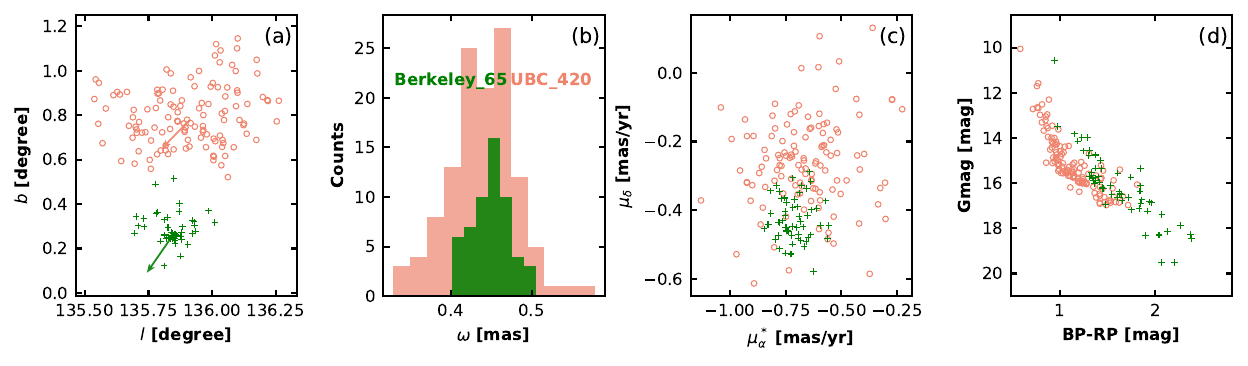}
    \includegraphics[width=160mm]{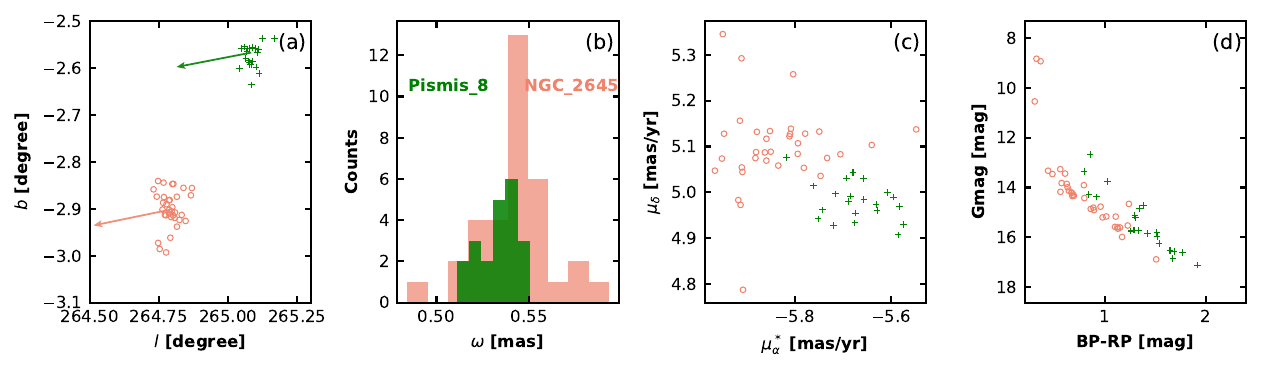}
    \includegraphics[width=160mm]{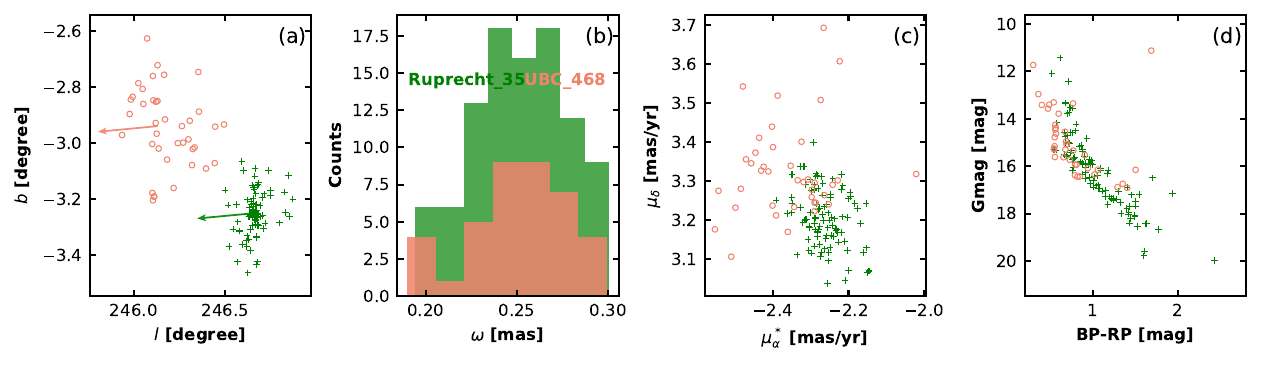}
      \caption{Same as Fig.~\ref{fig:the_same_cluster}, but for binary cluster candidates. The direction of the transverse motion of each component is marked by an arrow.}
         \label{fig:binary_pairs_1}
\end{figure*}

\citet{Palma2025} suggested that the separation between components of a bound binary cluster should be less than 50~pc, but this distance criterion alone is neither necessary nor sufficient to establish a gravitational bond. The question of whether one pair would be classified as binary or optical pair candidate is also dependent on a comprehensive analysis of its RV and TV. Orbit simulation is an additional tool to investigate the physical link between the two components of a binary system.

According to our criteria of 50~pc separation and similar transverse motion ($\Delta$TV~<~10~\kms and $\Delta$$\theta$~<~20$^{\degree}$), there are seven binary cluster candidates in our sample that are not part of a group, while 25 additional pairs form six cluster group candidates, making a total of 32 pairs from 14 independent CG20 clusters. The astrometric parameters of the two components in each pair are listed in Table~\ref{tab:pairs_para}, while their kinematic properties (including $\Delta$3D and TVs) are presented in Table~\ref{tab:velocity}, as well as the RV and age when available.

We performed orbit simulations on the seven isolated pairs (excluding NGC\_2645 versus Pismis\_8 and UBC\_468 versus Ruprecht\_35 due to the lack of RV) by employing the Galpy package and integrating the orbits forward for 500 Myr, as shown in Figs.~\ref{fig:Orbit_binary_pairs_1} and \ref{fig:Orbit_binary_pairs_2}. The simulations were achieved based on their coordinates, parallax, proper motions, and RV, with 500 MC samples, to take into account the uncertainties of the input parameters.

\paragraph{UBC\_207 versus NGC\_1980}
This pair is a strong candidate as a gravitationally bound binary cluster. The two clusters are separated by only $\Delta$3D = 10.61~pc, with remarkably consistent TVs ($\Delta$TV$\sim$0.3~\kms). As shown in Fig.~\ref{fig:binary_pairs_1}, the two clusters exhibit nearly identical proper motion vectors, while their parallax distributions and CMD overlap. Radial velocities are also in excellent agreement ($\Delta$RV$\sim$1.1~\kms). In addition, the ages of the two clusters are consistent, considering the typical age uncertainties of OC dating. The orbit simulation in Fig.~\ref{fig:Orbit_binary_pairs_1} reveals that their 3D separation remains relatively stable within 100~Myr. This suggests that this pair is likely gravitationally bound and originated from the same giant molecular cloud.

\paragraph{UBC\_420 versus Berkeley\_65} Although these two clusters have a larger separation than the previous pair, ($\Delta$3D = 48.13 pc), they have very consistent transverse motion and RV, as displayed in Table~\ref{tab:velocity}. It is worth to note that their RV values (i.e., $-$64.24~\kms\ and $-$60.43~\kms\, respectively) is based on only one member in each cluster, but agree within the typical error bars. Their tight clustering in velocity space and consistent CMD (see Fig.~\ref{fig:binary_pairs_1}), as well as the orbital stability revealed by the flat $\Delta$3D evolution curve (Fig.~\ref{fig:Orbit_binary_pairs_1}) indicate they likely formed from the same parental molecular cloud and have remained a bound system throughout their evolution.

\paragraph{NGC\_2645 versus Pismis\_8}
The two clusters exhibit remarkably consistent transverse motion ($\Delta$TV$\sim$0.7~\kms) and similar ages ($\Delta$log(Age)$\sim$0.26~dex), as shown in Fig.~\ref{fig:binary_pairs_1}, suggesting a common origin. However, the absence of RV measurements for Pismis\_8 in Table~\ref{tab:velocity} precludes any reliable orbital integration and definitive dynamical classification. Their kinematic coherence implies they may represent expanding fragments of a single star-forming complex, but their current binding status remains undetermined. Therefore, we classified this pair as a possible bound system based solely on its spatial and tangential kinematic properties, to be confirmed by RV measurements, which are currently lacking.

\paragraph{UBC\_468 versus Ruprecht\_35} This pair exhibits a three-dimensional separation of $\Delta$3D~=~49.25~pc, approaching our 50 pc threshold. As shown in Fig.~\ref{fig:binary_pairs_1}, the two clusters display distinct spatial distributions but comparable parallaxes and consistent transverse motion ($\Delta$TV$\sim$1.6~\kms). However, the lack of RV measurements for Ruprecht\_35 (Table~\ref{tab:velocity}), combined with UBC\_468's RV determination based on a single member star, precludes reliable orbital integration. Their ages show reasonable agreement ($\Delta$log(Age)$\sim$0.38~dex). While the spatial proximity and kinematic coherence favor a physical association, the incomplete velocity data render the gravitational binding status uncertain. We classify this pair as a possible binary system with a common origin based on consistent TV and overlapping CMDs, requiring spectroscopic follow-up to confirm radial velocities and establish physical association.

\paragraph{SAI\_72 versus CWNU\_1966}
The two clusters show coherent transverse motion direction but exhibit a substantial radial and TV differences ($\Delta$RV$\sim$11~\kms and $\Delta$TV$\sim$8~\kms), as presented in Table~\ref{tab:velocity}. While their spatial proximity in Fig.~\ref{fig:binary_pairs_2} suggests a possible physical association, the striking RV discrepancy and the large uncertainty in the H\&R24 value render the dynamical prediction highly uncertain. Despite these kinematic uncertainties, the simulated orbits of this pair in Fig.~\ref{fig:Orbit_binary_pairs_2} show periodic fluctuations in $\Delta$3D. This behavior may indicate a fly-by encounter or a loosely associated state. While there is no age information available for CG20 in Table~\ref{tab:velocity}, the CMD overlap in a way suggesting a similar age. The age quoted by H\&R24 is rather old, logAge~=~8.46, which is an additional argument against assumption of the two clusters being kinematically bound.

\paragraph{UBC\_14 versus ASCC\_106}
This pair displays comparable transverse motion ($\Delta$TV$\sim$1.7~\kms), but discrepant radial velocities ($\Delta$RV$\sim$9~\kms, Table~\ref{tab:velocity}). The orbital simulation reveals diverging trajectories with rapidly increasing 3D separation (Fig.~\ref{fig:Orbit_binary_pairs_2}), inconsistent with gravitational binding. We classified this pair as an optical alignment of unrelated clusters from the same star-forming region based on their similar age ($\Delta$log(Age)$\sim$0.37~dex) despite their spatial proximity.

\paragraph{UPK\_533 versus UPK\_535}
The two clusters also display diverging trajectories in the orbital simulation (Fig.~\ref{fig:Orbit_binary_pairs_2}) despite their spatial proximity, comparable transverse motion (Fig.~\ref{fig:binary_pairs_2}), and similar age. The significant discrepancy in RV ($\Delta$RV$\sim$23~\kms), as displayed in Table~\ref{tab:velocity}, combined with the expanding 3D separation over time, suggests they originated from a common molecular cloud, but are currently unbound and undergoing rapid dynamical separation.

It is important to note that the reliability of our kinematic classification varies significantly among these seven pairs due to the heterogeneous quality of RV measurements, with member star counts ranging from none to 93 (Table~\ref{tab:velocity}). For pairs with limited data, both the inferred gravitational binding and orbital predictions remain uncertain. Our conclusions rely primarily on transverse motion consistency and orbital stability. Thus, these findings should be regarded as tentative until confirmed by high-resolution spectroscopic observations of additional member stars.

\section{Cluster group candidates}\label{group}

The discovery of hierarchical structures in the Galactic disk, such as cluster groups and complexes, provides essential clues to the fragmentation of giant molecular clouds (GMCs) and the subsequent dynamical evolution of stellar systems. In this study, we identified seven clusters from CG20 forming pairs with several clusters from H\&R24, suggesting that they are part of groups. The parameters are given in Tables~\ref{tab:pairs_para} and \ref{tab:velocity} and the distribution of their members are represented in Figs.~\ref{fig: group_1} and~\ref{fig: group_2}. While the existence of these groups is generally documented in the literature \citep{Palma2025}, our multidimensional analysis based on Gaia DR3 data has led to the identification of new clusters in every group, thereby refining the spatial and kinematic census of these systems.

\paragraph{NGC\_1977 group} This group exhibits a remarkable kinematic coherence, with TV dispersion ($\Delta$TV$\sim$1.11~\kms) that is substantially lower than that of any other group in our sample. The RVs presented in Table~\ref{tab:velocity} confirm this exceptional association, with all members showing consistent values. The excellent agreement in both transverse and radial components, combined with their spatial concentration (Fig.~\ref{fig: group_1}), strongly argues against chance alignment and supports sequential star formation from a common parental molecular cloud. The age progression evident in the CMD and Table~\ref{tab:velocity} supports a scenario of multi-epoch star formation. \citet{Palma2025} reported this group as G10 with five candidate members. Our verification confirms three of these as kinematically coherent, but excludes OCSN\_61 and ASCC\_20 due to position angle differences exceeding $80^{\circ}$. We additionally identified four new members, expanding this to a seven-member group.

\paragraph{UBC\_392 group} This group comprises five clusters displaying coherent transverse motions with a velocity dispersion of merely $\sim$3.7~\kms and convergent parallax distributions ($\omega$$\sim$1.1~mas), as shown in Fig.~\ref{fig: group_1}. \citet{Palma2025} reported this group as G30 with seven candidate members. Our verification confirms four of these as kinematically coherent, but excludes Pismis\_Moreno\_1, HSC\_824, and FSR\_0398 due to position angle differences exceeding our $20^{\circ}$ threshold. However, the RVs reported in Table~\ref{tab:velocity} reveal a striking kinematical dichotomy between the central cluster and its proposed members, with differences exceeding 30~\kms. Combined with an age contrast of over 60~Myr, this may challenge a simple coeval origin and gravitational binding. We suggest this system may represent a dynamically assembled association or multigenerational star-forming complex, rather than a bound cluster group.

\paragraph{UBC\_672 + UBC\_553 group} Interestingly, the two groups formed by UBC\_672 and UBC\_553 in CG20 and the clusters in H\&R24, respectively, actually belong to the same cluster group, as shown in Fig.~\ref{fig: group_1}. This complex consists of six clusters centered on the massive cluster NGC\_6231. The high degree of consistency in parallax distributions and proper motion vectors in Fig.~\ref{fig: group_1} confirms that these clusters represent a coherent, comoving structure. The small 3D separations between several members (Table~\ref{tab:velocity}) support a deeply bound hierarchical structure. The RVs reported in Table~\ref{tab:velocity} show an excellent agreement among core members, although one cluster from H\&R24 lacks an RV value, while UBC\_672's RV relies on only one star. Besides, the ages of these clusters in Table~\ref{tab:velocity} are also consistent with a single star-forming event within the measurement uncertainties. \citet{Palma2025} reported this group as G44 with four member clusters, including CWNU\_1851 that is missing from our sample. However, we find that CWNU\_1851 exhibits a position angle differing by more than $60^{\circ}$ from the other members, exceeding our $20^{\circ}$ threshold. This means our inclusion of previously unassociated clusters significantly expands the known footprint of this complex.

\paragraph{FSR\_0198 group} This compact group comprises three clusters displaying consistent transverse motion ($\Delta$TV$\sim$5~\kms) and the same age (Table~\ref{tab:velocity}), as shown in Fig.~\ref{fig: group_1}. The RV data presented in Table~\ref{tab:velocity} do not agree. They are, however, very uncertain, relying on only one, two, or three stars, with a large dispersion. Two clusters (FSR\_0198 and Teutsch\_8) in this group were identified by \citet{Song2022} as a binary cluster candidate. However, this group has not been previously reported on in the literature \citep[e.g.][]{Casado2021, Palma2025, Liu2025}. Improved RV measurements are needed to definitively establish the dynamical state of this system.

\paragraph{Kronberger\_1 group} We found three clusters forming this group, with consistent transverse motion ($\Delta$TV$\sim$6~\kms) and age (Table~\ref{tab:velocity}). This system illustrates how center selection affects group membership definition. \citet{Palma2025} identified a group centered on HSC\_1351 including HSC\_1358 and Stock\_8 which are not paired to Kronberger\_1 according to our methodology. With Kronberger\_1 as the group center, HSC\_1358 and Stock\_8 fall outside our 50 pc spatial threshold. We verified that these clusters indeed satisfy our kinematic criteria, indicating that the group could include five clusters sharing consistent kinematics. This demonstrates that these clusters may form a kinematically coherent but spatially extended association, rather than a compact group satisfying strict proximity criteria. The lack of RV information in Table~\ref{tab:velocity} precludes a definitive dynamical assessment.

\paragraph{Bochum\_11 group} In this group, we identified four clusters exhibiting nearly similar transverse motion with velocity differences of merely $\sim$4~\kms and similar age ($\Delta$log(Age)$\sim$0.8~dex), as shown in Fig.~\ref{fig: group_2}. However, the RV data presented in Table~\ref{tab:velocity} reveal significant inconsistencies and large uncertainties, casting doubt on the physical association of the clusters despite their coherent transverse motion. \citet{Palma2025} reported this group as G21 with five candidate members including Gulliver\_52 that is missing in our sample. Our analysis confirms three members from their list but excludes Gulliver\_52 due to a TV discrepancy exceeding 20~\kms, and additionally identifies Bochum\_11 as a new member, resulting in a refined four-member system. High-resolution spectroscopic follow-up is essential to resolve the RV discrepancies and confirm the true membership.

Compared to recent large-scale surveys of cluster pairs and groups \citep[e.g.][]{Palma2025, Li2025}, our cross-match of CG20 and H\&R24 reveals that single-catalog studies yield incomplete membership censuses. The prevalence of new members suggests that molecular cloud fragmentation is more prolific than previously recorded, with many isolated clusters likely belonging to larger kinematic systems. This updated census provides a foundation for future spectroscopic studies of chemical homogeneity.

\section{Summary and conclusions}\label{summary}

This study presents a systematic search for missing multiple cluster systems by cross-matching the CG20 and H\&R24 catalogs. We identified seven new binary cluster candidates that had previously been overlooked, thanks to their proximity in space, as well as their similarity in velocity and age. We also identified seven clusters from CG20 that are new members of groups already reported in the literature. Our analysis emphasizes the necessity of multi-catalog integration and dynamical validation for establishing a complete census of the Galactic cluster population. Our primary conclusions are as follows:

\begin{itemize}

\item Our cross-matching identified 15 nomenclature discrepancies between CG20 and H\&R24, verified through overlapping distributions in position, proper motion, parallax, and CMDs. Four of them are newly identified duplicate objects, while others had previously been detected in the UCC.

\item Seven binary cluster candidates were identified based on their spatial proximity and kinematic consistency. Two pairs exhibit stable orbits over 100 Myr, identifying them as gravitational binding candidates. One pair shows diverging trajectories indicative of an unbound, dispersing system; however, its lack of RV data precludes definitive classification.

\item We expanded and refined seven established cluster groups. A systematic kinematic verification confirmed new members, while excluding previously proposed constituents that failed our selection criteria, substantially revising the spatial and mass distribution of these complexes.

\item Access to RV data is clearly essential for distinguishing true associations from chance alignments. The NGC\_1977 group demonstrates exceptional kinematic coherence, confirming its bound status; whereas the Bochum\_11 group exhibits RV inconsistencies that challenge membership verification. The heterogeneous quality of existing measurements represents a critical limitation, thereby necessitating a homogeneous spectroscopic follow-up.

\item The prevalence of new members suggests that molecular cloud fragmentation is more prolific than previously recorded, with many isolated clusters likely belonging to larger kinematic systems. Observed age gradients support sequential, multi-epoch star formation within fragmented giant molecular clouds.

\end{itemize}

In summary, these thirteen systems provide refined targets for future spectroscopic studies of chemical homogeneity, advancing our understanding of cluster formation in the Galactic disk. Moreover, by reconciling these inconsistencies that recovers hidden pairs overlooked by single-catalog studies, this work confirms that current binary cluster censuses remain incomplete. There is no doubt that the forthcoming Gaia DR4 will comprise a decisive step forward in the field.

\section{Data availability}

The table with 83 OCs from CG20 missing in H\&R24 and their new parameters based on Gaia DR3 is available at the CDS via \url{http://cdsarc.cds.unistra.fr/viz-bin/cat/J/AA/000/A000}, together with the table of binary and multiple OCs listed in Tables~\ref{tab:pairs_para} and \ref{tab:velocity}.

\begin{acknowledgements}
      We would like to thank the referee M.V. Kulesh for offering valuable suggestions that have enhanced the quality of this article. This work is supported by the National Natural Science Foundation of China (NSFC) under grant 12303037 and 12573035, the Natural Science Foundation of Sichuan Province (No.2025ZNSFSC0879), and the Fundamental Research Funds of China West Normal University of China (CWNU, No.23KE024). Qingshun Hu would like to acknowledge the financial support provided by the China Scholarship Council program (Grant No.202308510136). This work presents results from the European Space Agency (ESA) space mission Gaia. Gaia data are being processed by the Gaia Data Processing and Analysis Consortium (DPAC). Funding for the DPAC is provided by national institutions, in particular the institutions participating in the Gaia MultiLateral Agreement (MLA). The Gaia mission website is https://www.cosmos.esa.int/gaia. The Gaia archive website is https://archives.esac.esa.int/gaia. The preparation of this work has made extensive use of Topcat \citep{Taylor2005}, of \texttt{Astropy.Skycoord} package \citep{Astropy2013, Astropy2018}, of the Simbad and VizieR databases at CDS, Strasbourg, France, and of NASA's Astrophysics Data System Bibliographic Services.
      
\end{acknowledgements}

\bibliographystyle{aa}
\bibliography{references}

\onecolumn
\appendix

\section{Additional material}

\begin{figure*}[htbp]
   \centering
   \includegraphics[width=160mm]{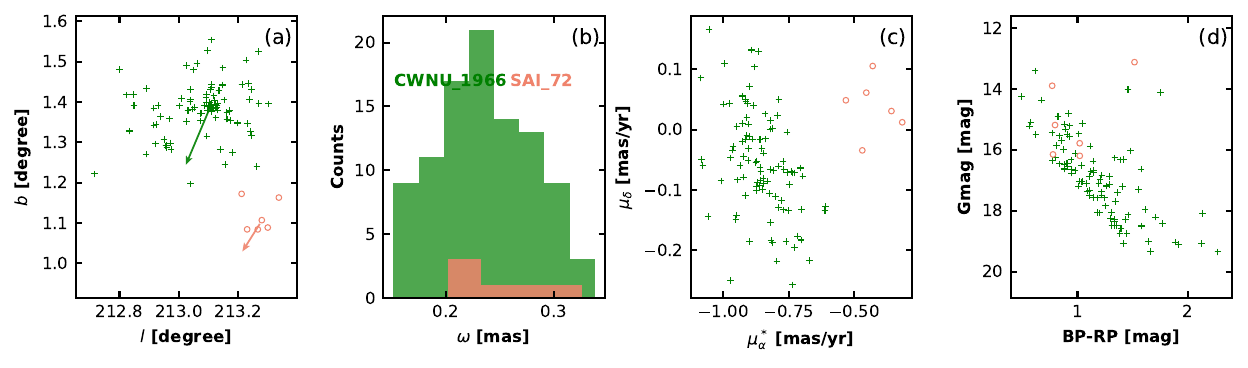}
    \includegraphics[width=160mm]{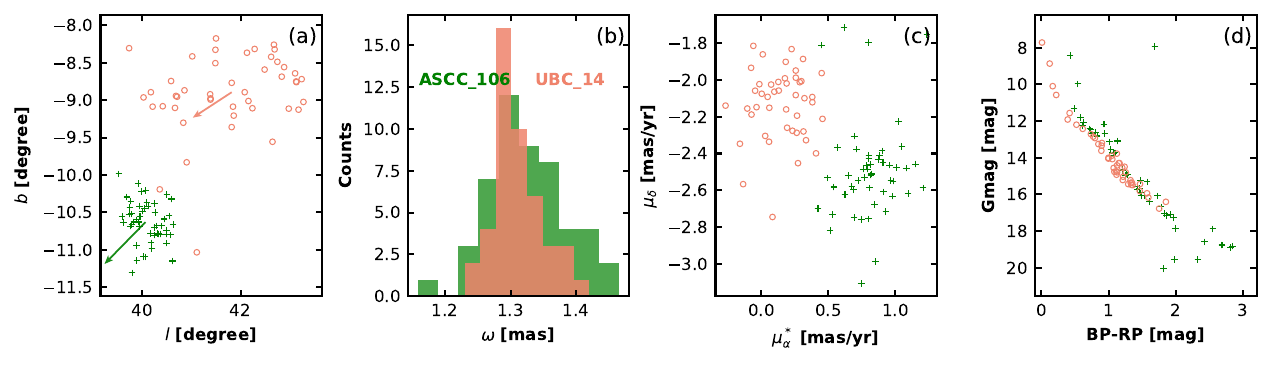}
    \includegraphics[width=160mm]{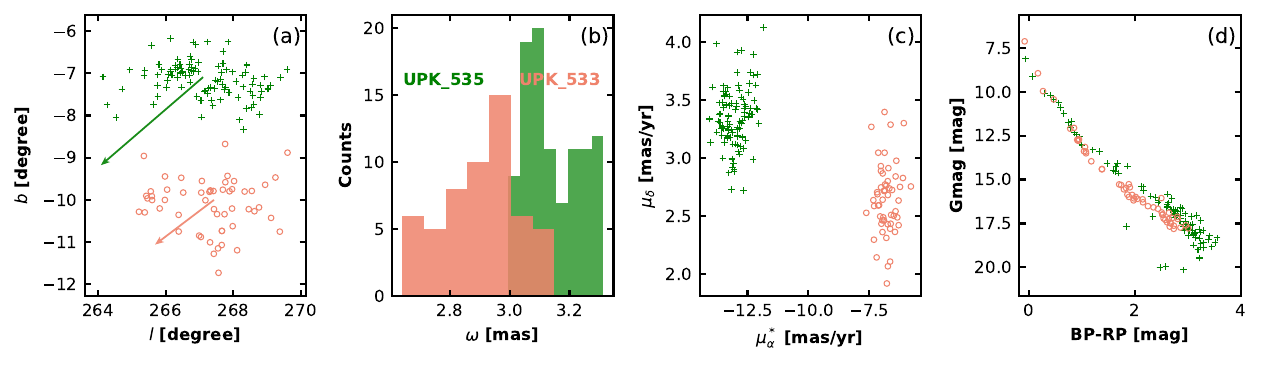}
      \caption{Continuation of Fig.\ref{fig:binary_pairs_1}}
         \label{fig:binary_pairs_2}
\end{figure*}

\begin{table*}
        \tiny
        \centering
        \begin{threeparttable}
        \caption{Astrometric parameters of cluster pair candidates.}
        \begin{tabular*}{\linewidth}{@{\extracolsep{\fill}}llcccccc}
                \toprule
                \toprule
                Cluster$_{\text{CG20}}$ & Cluster$_{\text{H\&R24}}$ & \pmra$_{\text{CG20}}$ & \pmra$_{\text{H\&R24}}$ & \pmdec$_{\text{CG20}}$ & \pmdec$_{\text{H\&R24}}$ & $\omega_{\text{CG20}}$ & $\omega_{\text{H\&R24}}$\\ 
                 {} & {} & \masyr & \masyr & \masyr & \masyr &  mas & mas \\
                \midrule
(1) & (2) & (3) & (4) & (5) & (6) & (7) &  (8) \\
                \midrule
UBC\_207 & NGC\_1980 & 1.01$\pm$0.18 & 1.26$\pm$0.01 & 0.59$\pm$0.29 & 0.47$\pm$0.01 & 2.57$\pm$0.03 & 2.60$\pm$0.002 \\
UBC\_420 & Berkeley\_65 & -0.66$\pm$0.10 & -0.71$\pm$0.01 & -0.26$\pm$0.08 & -0.43$\pm$0.01 & 0.44$\pm$0.03 & 0.45$\pm$0.003 \\
NGC\_2645 & Pismis\_8 & -5.86$\pm$0.06 & -5.67$\pm$0.01 & 5.09$\pm$0.04 & 4.98$\pm$0.01 & 0.54$\pm$0.01 & 0.53$\pm$0.002 \\
UBC\_468 & Ruprecht\_35 & -2.35$\pm$0.07 & -2.26$\pm$0.01 & 3.30$\pm$0.06 & 3.20$\pm$0.01 & 0.25$\pm$0.02 & 0.25$\pm$0.003 \\
SAI\_72 & CWNU\_1966 & -0.44$\pm$0.06 & -0.86$\pm$0.01 & 0.04$\pm$0.02 & -0.05$\pm$0.01 & 0.24$\pm$0.03 & 0.23$\pm$0.004 \\
UBC\_14 & ASCC\_106 & 0.18$\pm$0.13 & 0.81$\pm$0.03 & -2.09$\pm$0.10 & -2.49$\pm$0.04 & 1.31$\pm$0.02 & 1.33$\pm$0.008 \\
UPK\_533 & UPK\_535 & -6.86$\pm$0.21 & -13.02$\pm$0.05 & 2.64$\pm$0.18 & 3.35$\pm$0.02 & 2.92$\pm$0.07 & 3.15$\pm$0.010 \\
 &  &  &  &  & & & \\
\midrule
 &  &  &  &  & & & \\
NGC\_1977 & ASCC\_19 & 1.28$\pm$0.31 & 1.17$\pm$0.02 & -0.73$\pm$0.33 & -1.18$\pm$0.02 & 2.54$\pm$0.03 & 2.81$\pm$0.004 \\
NGC\_1977 & HSC\_1633 & 1.28$\pm$0.31 & 1.84$\pm$0.02 & -0.73$\pm$0.33 & -1.04$\pm$0.02 & 2.54$\pm$0.03 & 2.63$\pm$0.005 \\
NGC\_1977 & OC\_0339 & 1.28$\pm$0.31 & 1.67$\pm$0.02 & -0.73$\pm$0.33 & -1.03$\pm$0.03 & 2.54$\pm$0.03 & 2.76$\pm$0.003 \\
NGC\_1977 & OCSN\_59 & 1.28$\pm$0.31 & 1.20$\pm$0.03 & -0.73$\pm$0.33 & -0.86$\pm$0.03 & 2.54$\pm$0.03 & 2.58$\pm$0.005 \\
NGC\_1977 & Sigma\_Orionis & 1.28$\pm$0.31 & 1.48$\pm$0.03 & -0.73$\pm$0.33 & -0.61$\pm$0.03 & 2.54$\pm$0.03 & 2.50$\pm$0.005 \\
\smallskip
NGC\_1977 & UBC\_17a & 1.28$\pm$0.31 & 1.80$\pm$0.02 & -0.73$\pm$0.33 & -1.30$\pm$0.02 & 2.54$\pm$0.03 & 2.77$\pm$0.005 \\
UBC\_392 & CWNU\_446 & -0.65$\pm$0.08 & -0.33$\pm$0.02 & -3.37$\pm$0.06 & -2.55$\pm$0.03 & 1.06$\pm$0.01 & 1.02$\pm$0.012 \\
UBC\_392 & CWNU\_1126 & -0.65$\pm$0.08 & -1.41$\pm$0.02 & -3.37$\pm$0.06 & -3.25$\pm$0.02 & 1.06$\pm$0.01 & 1.08$\pm$0.008 \\
UBC\_392 & CWNU\_1228 & -0.65$\pm$0.08 & -0.47$\pm$0.03 & -3.37$\pm$0.06 & -3.03$\pm$0.02 & 1.06$\pm$0.01 & 1.08$\pm$0.011 \\
\smallskip
UBC\_392 & OC\_0185 & -0.65$\pm$0.08 & -1.30$\pm$0.02 & -3.37$\pm$0.06 & -2.49$\pm$0.02 & 1.06$\pm$0.01 & 1.08$\pm$0.003 \\
UBC\_672 & HSC\_2848 & -0.67$\pm$0.03 & -0.58$\pm$0.02 & -2.02$\pm$0.05 & -1.70$\pm$0.02 & 0.61$\pm$0.03 & 0.62$\pm$0.002 \\
UBC\_672 & HSC\_2849 & -0.67$\pm$0.03 & -0.46$\pm$0.02 & -2.02$\pm$0.05 & -1.59$\pm$0.02 & 0.61$\pm$0.03 & 0.60$\pm$0.002 \\
UBC\_672 & NGC\_6231 & -0.67$\pm$0.03 & -0.58$\pm$0.01 & -2.02$\pm$0.05 & -2.18$\pm$0.01 & 0.61$\pm$0.03 & 0.61$\pm$0.002 \\
\smallskip
UBC\_672 & OC\_0673 & -0.67$\pm$0.03 & -0.91$\pm$0.01 & -2.02$\pm$0.05 & -2.01$\pm$0.01 & 0.61$\pm$0.03 & 0.61$\pm$0.003 \\
UBC\_553 & ESO\_332-13 & -0.56$\pm$0.08 & -0.16$\pm$0.01 & -1.86$\pm$0.10 & -1.14$\pm$0.02 & 0.60$\pm$0.02 & 0.59$\pm$0.002 \\
UBC\_553 & HSC\_2849 & -0.56$\pm$0.08 & -0.46$\pm$0.02 & -1.86$\pm$0.10 & -1.59$\pm$0.02 & 0.60$\pm$0.02 & 0.60$\pm$0.002 \\
UBC\_553 & NGC\_6231 & -0.56$\pm$0.08 & -0.58$\pm$0.01 & -1.86$\pm$0.10 & -2.18$\pm$0.01 & 0.60$\pm$0.02 & 0.61$\pm$0.002 \\
\smallskip
UBC\_553 & OC\_0673 & -0.56$\pm$0.08 & -0.91$\pm$0.01 & -1.86$\pm$0.10 & -2.01$\pm$0.01 & 0.60$\pm$0.02 & 0.61$\pm$0.003 \\
FSR\_0198 & Casado\_82 & -3.56$\pm$0.15 & -3.20$\pm$0.01 & -6.62$\pm$0.13 & -6.26$\pm$0.01 & 0.49$\pm$0.04 & 0.49$\pm$0.003 \\
\smallskip
FSR\_0198 & Teutsch\_8 & -3.56$\pm$0.15 & -3.37$\pm$0.01 & -6.62$\pm$0.13 & -6.60$\pm$0.01 & 0.49$\pm$0.04 & 0.50$\pm$0.003 \\
Kronberger\_1 & HSC\_1351 & -0.06$\pm$0.06 & 0.12$\pm$0.02 & -2.33$\pm$0.09 & -1.74$\pm$0.02 & 0.46$\pm$0.04 & 0.46$\pm$0.005 \\
\smallskip
Kronberger\_1 & HSC\_1356 & -0.06$\pm$0.06 & 0.16$\pm$0.01 & -2.33$\pm$0.09 & -1.87$\pm$0.01 & 0.46$\pm$0.04 & 0.46$\pm$0.004 \\
Bochum\_11 & ASCC\_62 & -6.34$\pm$0.07 & -6.39$\pm$0.01 & 1.79$\pm$0.09 & 2.09$\pm$0.01 & 0.40$\pm$0.02 & 0.40$\pm$0.003 \\
Bochum\_11 & Collinder\_228 & -6.34$\pm$0.07 & -6.67$\pm$0.02 & 1.79$\pm$0.09 & 1.92$\pm$0.01 & 0.40$\pm$0.02 & 0.40$\pm$0.002 \\
Bochum\_11 & Trumpler\_16 & -6.34$\pm$0.07 & -6.91$\pm$0.01 & 1.79$\pm$0.09 & 2.61$\pm$0.01 & 0.40$\pm$0.02 & 0.40$\pm$0.002 \\
\bottomrule
        \end{tabular*}
        
        \label{tab:pairs_para}
    \end{threeparttable}
\end{table*}

\begin{table*}
        \tiny
        \centering
        \caption{Kinematic and physical parameters of cluster pair candidates.}
        
        \begin{tabular*}{\textwidth}{@{\extracolsep{\fill}}llccccccccc}
        \toprule
        \toprule
        Cluster$_{\text{CG20}}$ & Cluster$_{\text{H\&R24}}$ & $\Delta$3D & RV$_{\text{CG20}}$ & N$_{\text{CG20}}$ & RV$_{\text{H\&R24}}$ & N$_{\text{H\&R24}}$ & TV$_{\text{CG20}}$ & TV$_{\text{H\&R24}}$ & logAge$_{\text{CG20}}$ &   logAge$_{\text{H\&R24}}$\\ 
        {} & {} & pc & \kms & {} & \kms & {} & \kms & \kms & dex & dex \\
        \midrule
        (1) & (2) & (3) & (4) & (5) & (6) & (7) &  (8) & (9) & (10) &  (11) \\
                \midrule
UBC\_207 & NGC\_1980 & 10.61 & 22.12$\pm$10.62 & 24 & 23.24$\pm$2.84 & 93 & 2.17$\pm$0.25 & 2.46$\pm$0.01 & 7.18 & 6.84$_{6.63}^{6.99}$ \\
UBC\_420 & Berkeley\_65 & 48.13 & -64.24$\pm$1.62 & 1 & -60.43$\pm$7.02 & 1 & 7.64$\pm$0.76 & 8.78$\pm$0.08 & 7.81 & 7.24$_{7.07}^{7.44}$ \\
NGC\_2645 & Pismis\_8 & 32.93 & -25.27$\pm$4.52 & 2 & - & 0 & 67.75$\pm$1.03 & 67.06$\pm$0.22 & 7.41 & 7.67$_{7.42}^{7.96}$ \\
UBC\_468 & Ruprecht\_35 & 49.25 & 44.05$\pm$0.23 & 1 & - & 0 & 75.50$\pm$3.75 & 73.87$\pm$0.55 & 7.04 & 7.42$_{7.19}^{7.58}$ \\
SAI\_72 & CWNU\_1966 & 37.41 & 44.56$\pm$1.26 & 1 & 33.76$\pm$21.68 & 2 & 9.04$\pm$0.91 & 17.43$\pm$0.26 & - & 8.46$_{8.26}^{8.69}$ \\
UBC\_14 & ASCC\_106 & 34.10 & -6.28$\pm$7.45 & 22 & -15.38$\pm$1.81 & 20 & 7.71$\pm$0.28 & 9.39$\pm$0.11 & 7.72 & 8.09$_{7.87}^{8.27}$ \\
UPK\_533 & UPK\_535 & 31.26 & 27.12$\pm$3.70 & 17 & 3.91$\pm$4.45 & 25 & 11.91$\pm$0.33 & 20.24$\pm$0.06 & 7.95 & 7.34$_{7.06}^{7.57}$ \\
&  &  &  &  & & & & & &  \\
\midrule
&  &  &  &  & & & & & &  \\
NGC\_1977 & ASCC\_19 & 43.86 & 25.38$\pm$7.68 & 31 & 11.65$\pm$3.37 & 26 & 2.84$\pm$0.42 & 2.81$\pm$0.03 & 7.99 & 6.87$_{6.67}^{7.04}$ \\
NGC\_1977 & HSC\_1633 & 25.16 & 25.38$\pm$7.68 & 31 & 14.80$\pm$5.66 & 17 & 2.84$\pm$0.42 & 3.82$\pm$0.03 & 7.99 & 6.78$_{6.55}^{6.94}$ \\
NGC\_1977 & OC\_0339 & 43.00 & 25.38$\pm$7.68 & 31 & 29.10$\pm$7.32 & 3 & 2.84$\pm$0.42 & 3.37$\pm$0.03 & 7.99 & 7.90$_{6.79}^{8.50}$ \\
NGC\_1977 & OCSN\_59 & 48.91 & 25.38$\pm$7.68 & 31 & 17.39$\pm$4.30 & 13 & 2.84$\pm$0.42 & 2.71$\pm$0.04 & 7.99 & 6.77$_{6.51}^{6.90}$ \\
NGC\_1977 & Sigma\_Orionis & 17.53 & 25.38$\pm$7.68 & 31 & 24.97$\pm$4.15 & 31 & 2.84$\pm$0.42 & 3.05$\pm$0.05 & 7.99 & 6.57$_{6.43}^{6.65}$ \\
\smallskip
NGC\_1977 & UBC\_17a & 38.63 & 25.38$\pm$7.68 & 31 & 11.74$\pm$3.44 & 35 & 2.84$\pm$0.42 & 3.80$\pm$0.02 & 7.99 & 6.66$_{6.48}^{6.76}$ \\
UBC\_392 & CWNU\_446 & 43.78 & -3.26$\pm$3.12 & 7 & -33.23$\pm$7.82 & 6 & 15.38$\pm$0.23 & 11.88$\pm$0.15 & 7.85 & 6.57$_{6.42}^{6.66}$ \\
UBC\_392 & CWNU\_1126 & 33.10 & -3.26$\pm$3.12 & 7 & -25.15$\pm$2.73 & 17 & 15.38$\pm$0.23 & 15.60$\pm$0.11 & 7.85 & 7.20$_{6.96}^{7.40}$ \\
UBC\_392 & CWNU\_1228 & 27.59 & -3.26$\pm$3.12 & 7 & -38.19$\pm$6.19 & 1 & 15.38$\pm$0.23 & 13.42$\pm$0.13 & 7.85 & 6.56$_{6.40}^{6.65}$ \\
\smallskip
UBC\_392 & OC\_0185 & 37.78 & -3.26$\pm$3.12 & 7 & -20.23$\pm$3.23 & 54 & 15.38$\pm$0.23 & 12.37$\pm$0.06 & 7.85 & 6.81$_{6.64}^{6.94}$ \\
UBC\_672 & HSC\_2848 & 31.60 & -34.47$\pm$5.48 & 1 & - & 0 & 16.46$\pm$0.70 & 13.80$\pm$0.11 & 7.02 & 7.05$_{6.81}^{7.27}$ \\
UBC\_672 & HSC\_2849 & 30.86 & -34.47$\pm$5.48 & 1 & -30.33$\pm$14.88 & 3 & 16.46$\pm$0.70 & 13.06$\pm$0.10 & 7.02 & 6.61$_{6.46}^{6.70}$ \\
UBC\_672 & NGC\_6231 & 13.23 & -34.47$\pm$5.48 & 1 & -30.80$\pm$4.22 & 57 & 16.46$\pm$0.70 & 17.59$\pm$0.05 & 7.02 & 6.64$_{6.51}^{6.73}$ \\
\smallskip
UBC\_672 & OC\_0673 & 13.89 & -34.47$\pm$5.48 & 1 & -34.28$\pm$5.41 & 6 & 16.46$\pm$0.70 & 17.07$\pm$0.08 & 7.02 & 6.86$_{6.66}^{7.04}$ \\
UBC\_553 & ESO\_332-13 & 31.86 & 14.91$\pm$42.96 & 2 & -18.32$\pm$6.49 & 18 & 15.44$\pm$0.64 & 9.27$\pm$0.09 & 7.12 & 6.62$_{6.49}^{6.69}$ \\
UBC\_553 & HSC\_2849 & 10.16 & 14.91$\pm$42.96 & 2 & -30.33$\pm$14.88 & 3 & 15.44$\pm$0.64 & 13.06$\pm$0.10 & 7.12 & 6.61$_{6.46}^{6.70}$ \\
UBC\_553 & NGC\_6231 & 30.95 & 14.91$\pm$42.96 & 2 & -30.80$\pm$4.22 & 57 & 15.44$\pm$0.64 & 17.59$\pm$0.05 & 7.12 & 6.64$_{6.51}^{6.73}$ \\
\smallskip
UBC\_553 & OC\_0673 & 42.80 & 14.91$\pm$42.96 & 2 & -34.28$\pm$5.41 & 6 & 15.44$\pm$0.64 & 17.07$\pm$0.08 & 7.12 & 6.86$_{6.66}^{7.04}$ \\
FSR\_0198 & Casado\_82 & 30.43 & -9.55$\pm$4.46 & 2 & -22.38$\pm$8.49 & 1 & 72.40$\pm$3.69 & 67.34$\pm$0.31 & 6.67 & 6.77$_{6.59}^{6.90}$ \\
\smallskip
FSR\_0198 & Teutsch\_8 & 43.79 & -9.55$\pm$4.46 & 2 & 11.42$\pm$14.31 & 3 & 72.40$\pm$3.69 & 70.50$\pm$0.27 & 6.67 & 6.63$_{6.46}^{6.71}$ \\
Kronberger\_1 & HSC\_1351 & 15.90 & - & 0 & - & 0 & 23.93$\pm$1.65 & 17.91$\pm$0.24 & 6.78 & 6.80$_{6.58}^{6.98}$ \\
\smallskip
Kronberger\_1 & HSC\_1356 & 16.99 & - & 0 & 79.41$\pm$8.86 & 1 & 23.93$\pm$1.65 & 19.37$\pm$0.15 & 6.78 & 7.14$_{6.90}^{7.35}$ \\
Bochum\_11 & ASCC\_62 & 44.52 & -24.73$\pm$60.56 & 7 & 41.38$\pm$20.97 & 1 & 78.36$\pm$2.94 & 78.84$\pm$0.35 & 6.80 & 7.53$_{7.32}^{7.73}$ \\
Bochum\_11 & Collinder\_228 & 18.05 & -24.73$\pm$60.56 & 7 & -28.41$\pm$14.03 & 2 & 78.36$\pm$2.94 & 82.39$\pm$0.34 & 6.80 & 6.72$_{6.55}^{6.85}$ \\
Bochum\_11 & Trumpler\_16 & 32.74 & -24.73$\pm$60.56 & 7 & -2.88$\pm$13.41 & 14 & 78.36$\pm$2.94 & 87.02$\pm$0.24 & 6.80 & 6.79$_{6.58}^{7.00}$ \\
        \bottomrule
        
        \end{tabular*}
        \tablefoot{Columns (1) and (2) give the cluster names from CG20 and H\&R24, respectively. Column (3) gives the three‑dimensional separation ($\Delta 3\mathrm{D}$) between the two clusters in each pair. Columns (4) and (5) report the RV of the CG20 cluster and the number of members used in its calculation. Columns (6) and (7) list the RV of the H\&R24 cluster (taken directly from the catalog) and the corresponding number of members with RV measurements. Columns (8) and (9) provide the TVs of the clusters. Columns (10) and (11) present the ages of the clusters.}
          \label{tab:velocity}
\end{table*}

\begin{figure*}[htbp]
   \centering
   \includegraphics[width=160mm]{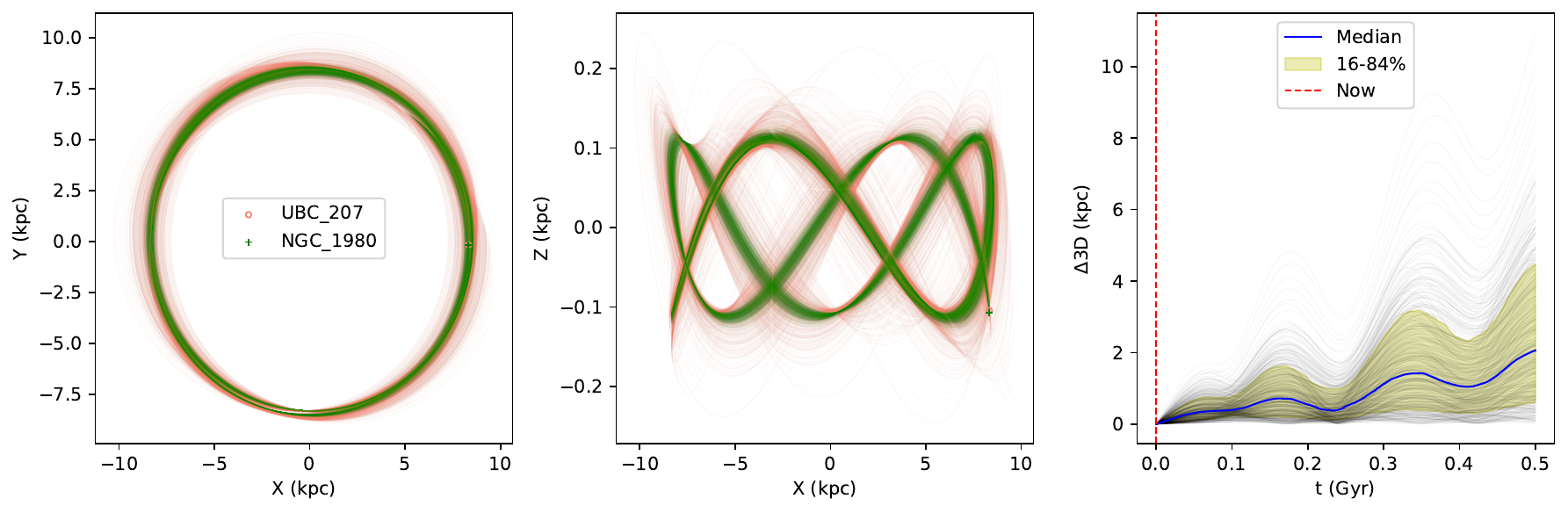}
   \includegraphics[width=160mm]{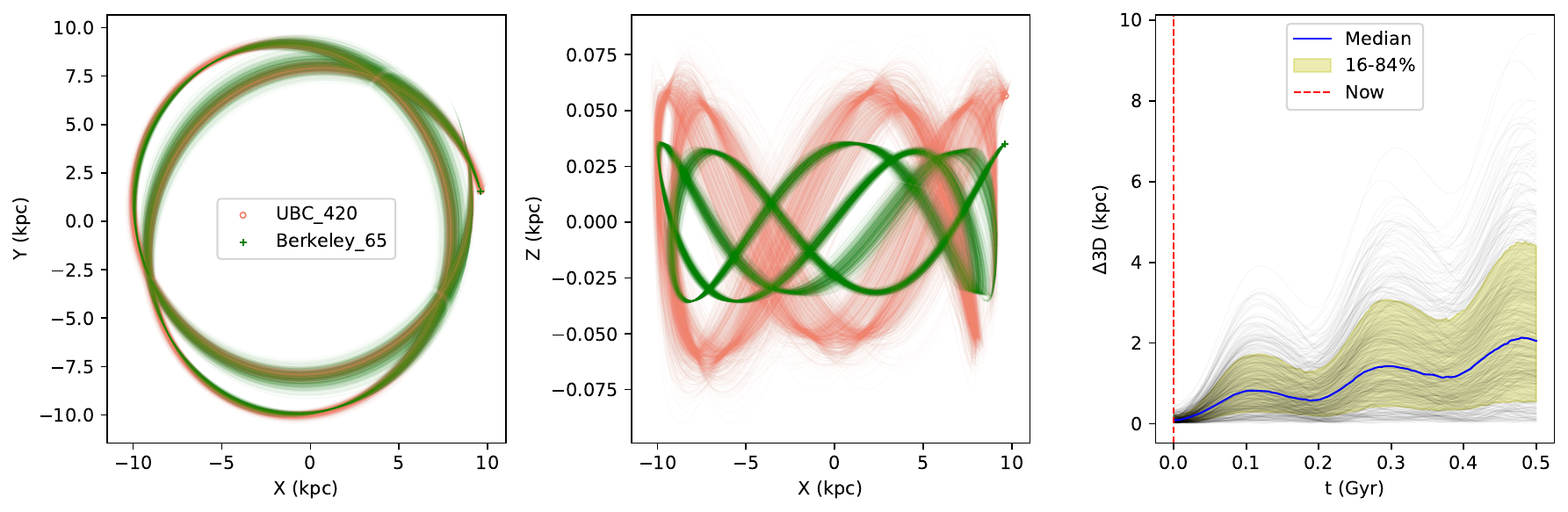}
      \caption{The orbit distributions of the binary cluster candidates within 500~Myr, as well as the trends of change in $\Delta$3D.}
         \label{fig:Orbit_binary_pairs_1}
\end{figure*}

\begin{figure*}[htbp]
   \centering
   \includegraphics[width=160mm]{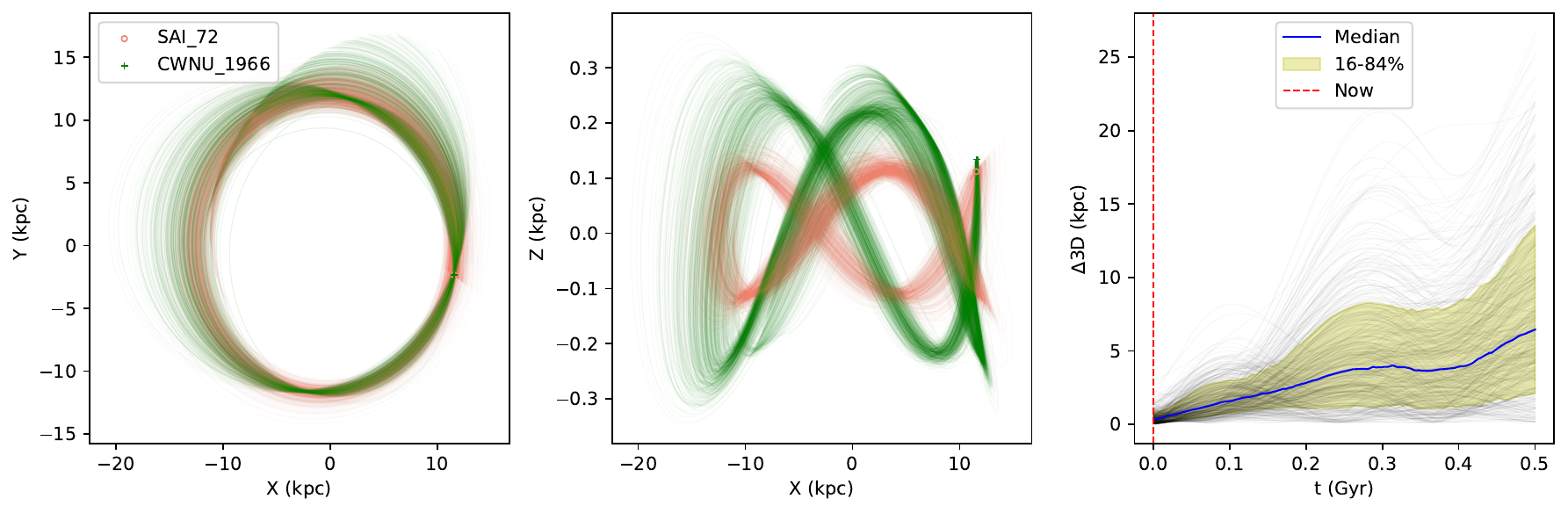}
    \includegraphics[width=160mm]{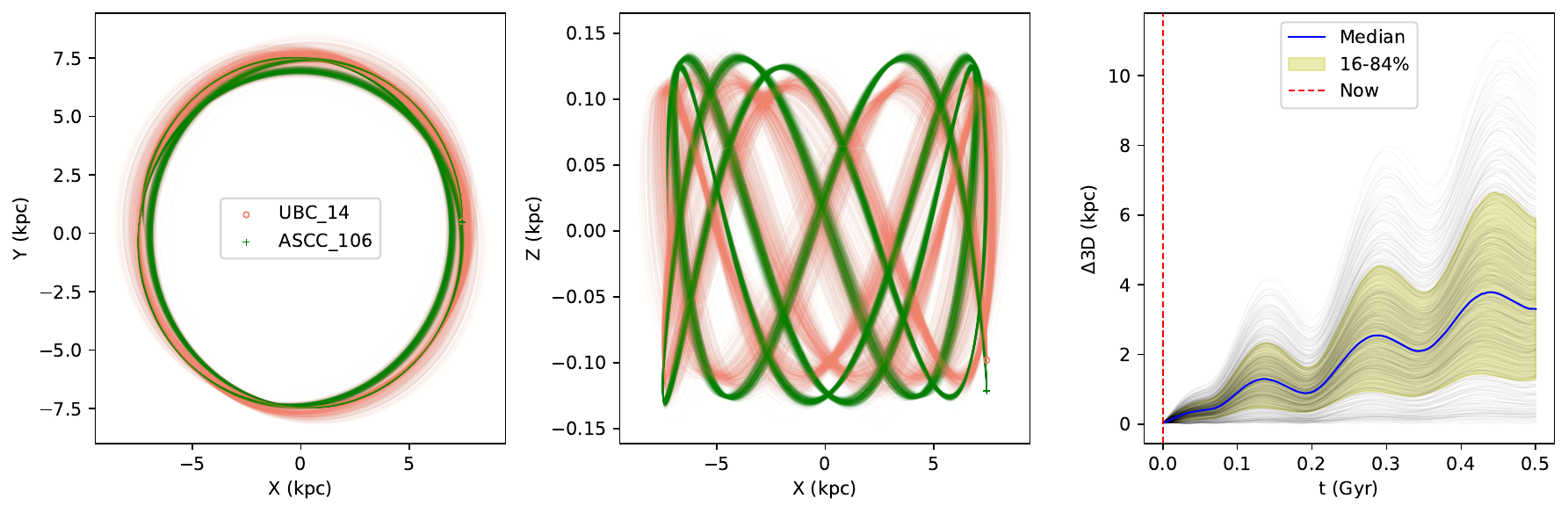}
    \includegraphics[width=160mm]{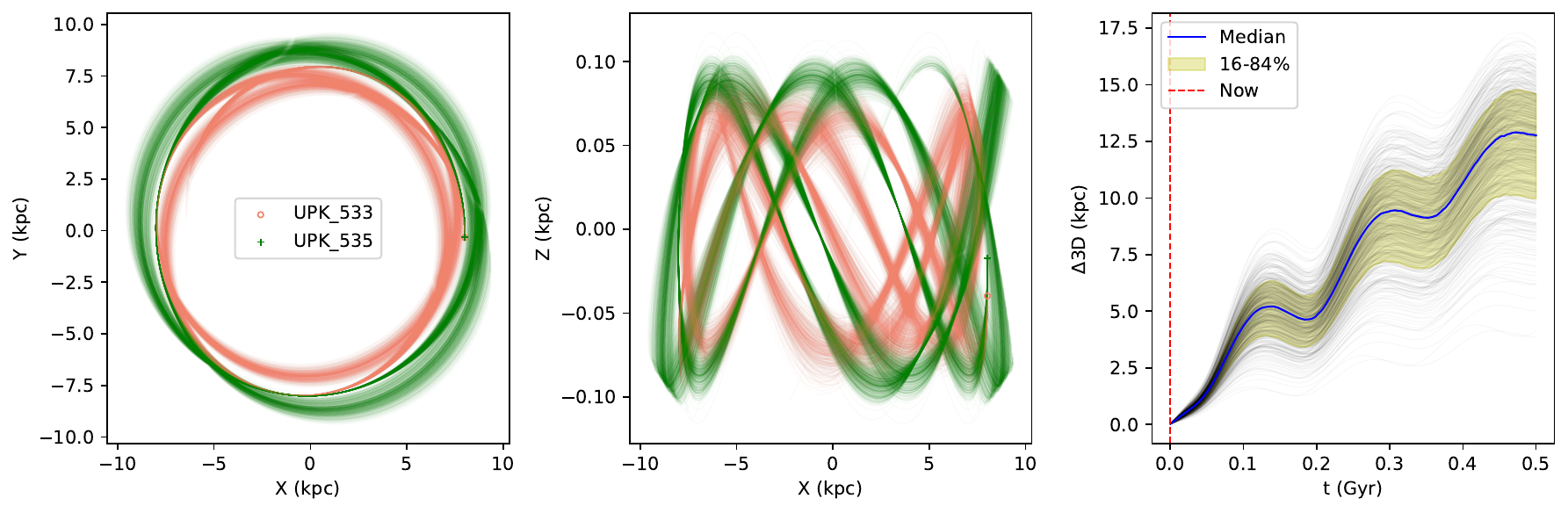}
      \caption{Continuation of Fig. \ref{fig:Orbit_binary_pairs_1}}
         \label{fig:Orbit_binary_pairs_2}
\end{figure*}

\begin{figure*}
   \centering
   \includegraphics[width=160mm]{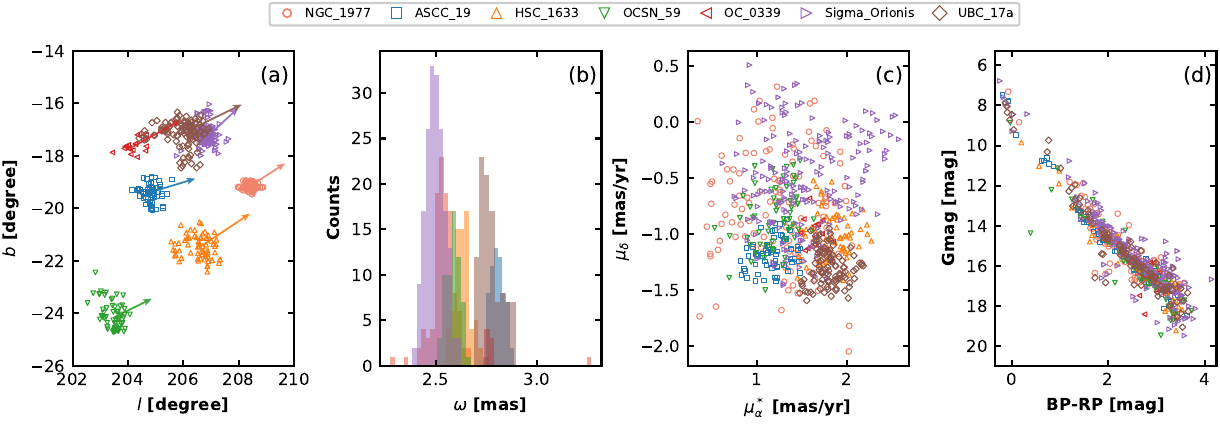}
   \includegraphics[width=160mm]{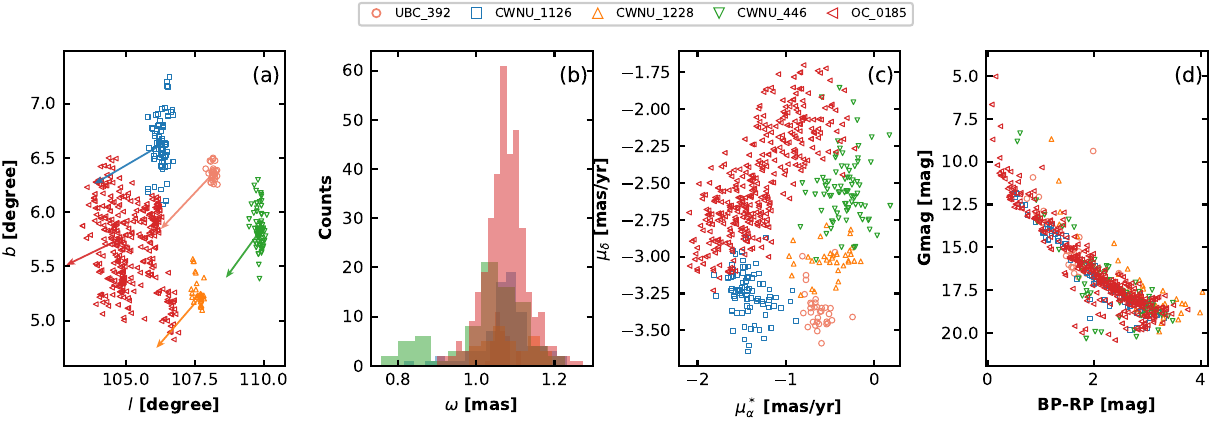}
   \includegraphics[width=160mm]{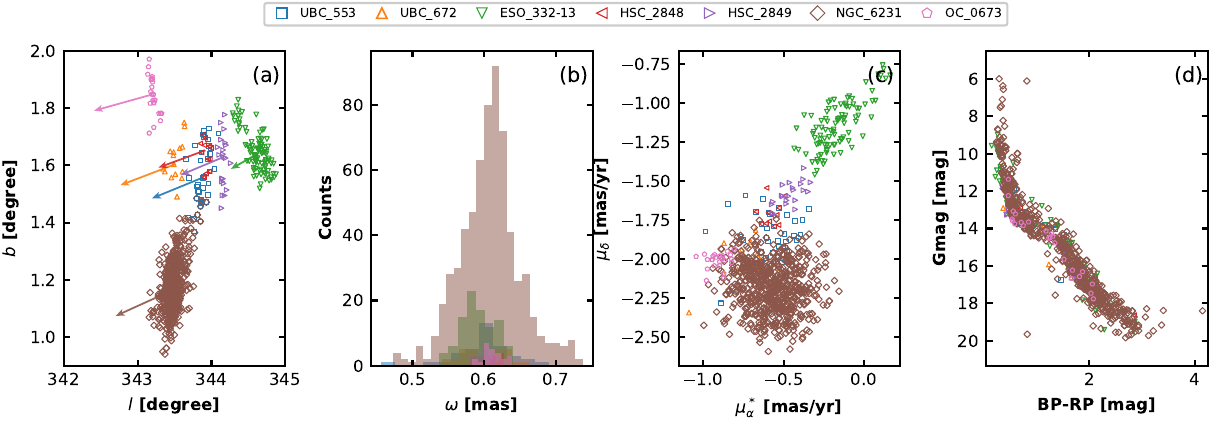}
   \includegraphics[width=160mm]{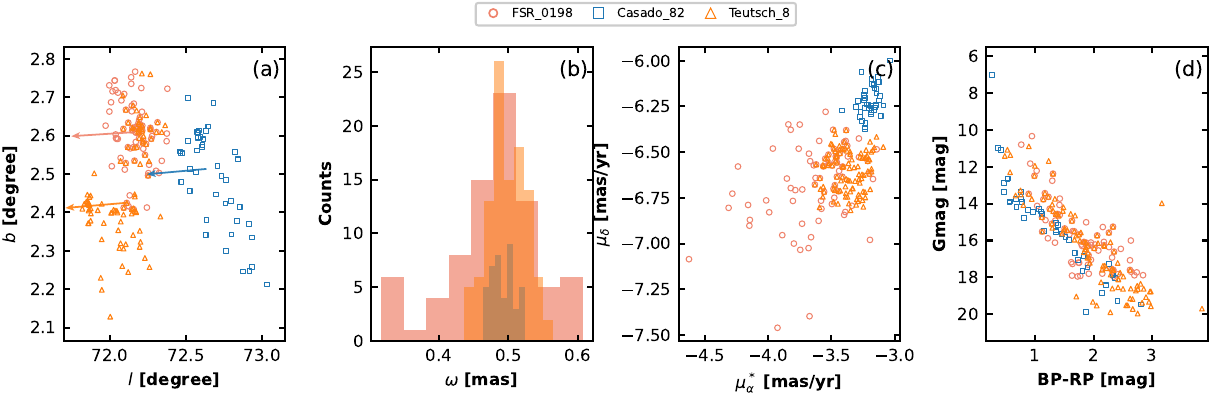}
    \caption{Same as Fig.~\ref{fig:binary_pairs_1}, but cluster group candidates.}
    \label{fig: group_1}
\end{figure*}

\begin{figure*}
   \centering
    \includegraphics[width=160mm]{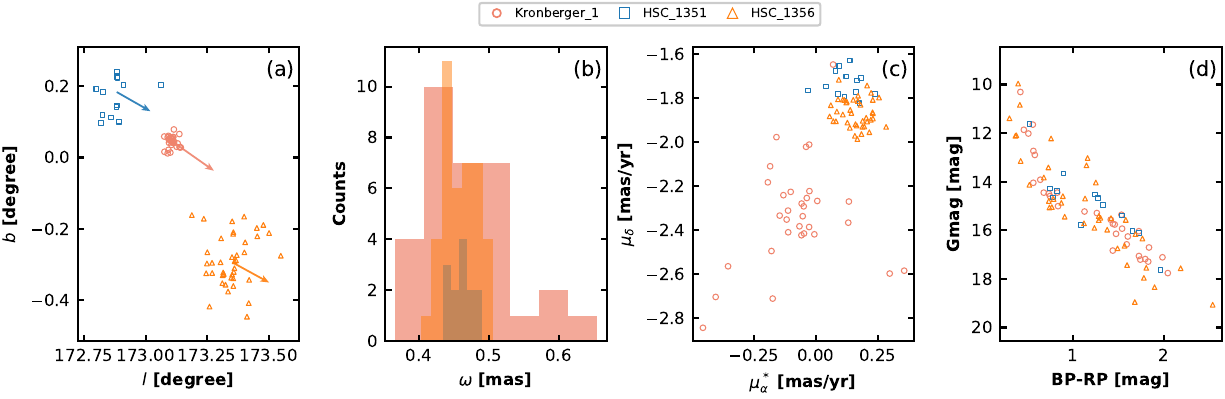}
    \includegraphics[width=160mm]{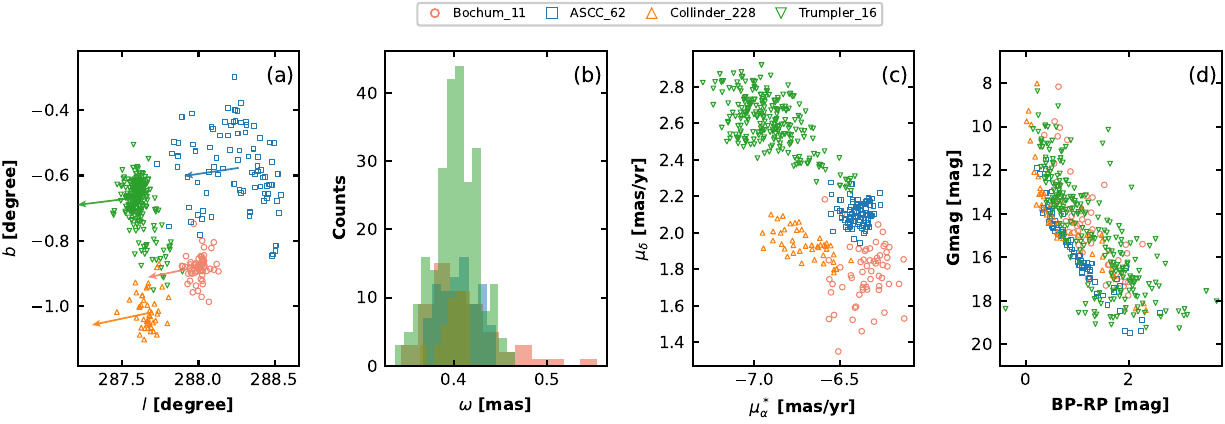}
    \caption{Continuation of Fig. \ref{fig: group_1}}
    \label{fig: group_2}
\end{figure*}

\end{document}